\documentclass[aps,prb,superscriptaddress,twocolumn]{revtex4-1}
\usepackage{graphicx}
\usepackage{amsmath}
\usepackage[english]{babel}
\usepackage{array}
\usepackage{amssymb, bm}

\begin{document}
\title{Spin splitting of two dimensional states in the conduction band of asymmetric heterostructures: contribution from the atomically sharp interface}

\author{Zh.A. Devizorova}
\email{DevizorovaZhanna@gmail.com}
\affiliation
{Moscow Institute of Physics and Technology, 141700 Dolgoprudnyi, Moscow District, Russia}
\affiliation
{V.A. Kotelnikov Institute of Radio Engineering and Electronics, Russian Academy of Sciences, 125009 Moscow, Russia}

\author{V.A.Volkov}
\email{Volkov.V.A.@gmail.com}
\affiliation
{V.A. Kotelnikov Institute of Radio Engineering and Electronics, Russian Academy of Sciences, 125009 Moscow, Russia}
\affiliation
{Moscow Institute of Physics and Technology, 141700 Dolgoprudnyi, Moscow District, Russia}

\begin{abstract}
The effect of an atomically sharp impenetrable interface on the spin splitting of the spectrum of two-dimensional electrons in heterostructures  based on (001) III-V compounds has been analyzed. To this end, the single band Hamiltonian  $\Gamma_{6c}$ for envelope functions is supplemented by a general boundary condition taking into account the possibility of the existence of Tamm states. This boundary condition also takes into account the spin-orbit interaction,  the asymmetry of a quantum well, and the lack of inversion symmetry in the crystal and contains the single phenomenological length $R$ characterizing the structure of the interface at atomic scales. The model of a quasitriangular well created by the electric field $F$ has been considered. After the unitary transformation to zero boundary conditions, in the modified Hamiltonian interfacial contribution appears, from which the two-dimensional spin Hamiltonian is obtained through averaging over the fast motion along the normal. In the absence of magnetic field $\boldsymbol B$, this contribution is the sum of the Dresselhaus and the Bychkov-Rashba terms with the constants renormalized owing to the interfacial contribution. In the field $\boldsymbol B$ containing the quantizing component $B_z$, the off - diagonal (in cubic axes) components of the $g$-factor tensor are linear functions of $|B_z|$ and the number of the Landau level $N$. The results are in qualitative agreement with the experimental data.
\end{abstract}
\maketitle

\section{Introduction.}
The spin-orbit interaction results in the spin splitting of the energies of two-dimensional electrons in asymmetric structures based on III-V compounds. The interaction with electric fields is described by various spin-dependent contributions to the effective two-dimensional Hamiltonian:

\begin{equation}
\label{2D ham}
\Delta \hat H_{2D}=\alpha_{BIA} (\sigma_y p_y - \sigma_x p_x) + \alpha _{SIA} (\sigma_x p_y -\sigma_y p_x).
\end{equation}
Here, $x$, $y$, and $z$ are the cubic axes, where $z$ is the quantum confinement axis; and $\sigma _x$,$\sigma _y$, and $\sigma _z$ are the Pauli matrices. The first term in Eq. (\ref{2D ham}) is due to lack of inversion symmetry in the bulk crystal potential (the Dresselhaus term \cite{s1, s2, s3, s4, s5}). The second term (the Bychkov-Rashba term \cite{s6, s7}, known also as the Rashba interaction) is associated with the asymmetry of the potential $V(z)$ of the structure.

The method of effective wavefunctions, which are envelopes of the total wavefunction, is most often used to derive the explicit expressions for the Dresselhaus, $\alpha _{BIA}$, and the Rashba, $\alpha _{SIA}$, constants. In this approximation, these constants are usually given by the expressions \cite{s1, s2, s3, s4, s5} 
\begin{equation}
\label{const BIA SIA}
\alpha_{BIA}^{(0)} = \frac{\gamma_c (\hat p_z ^2)_{00} }{\hbar ^3}, \qquad \alpha _{SIA}^{(0)} =  a_{SO} (\partial_z V)_{00},
\end{equation}
where $\gamma_c$ is the constant of the spin splitting of the conduction band of a III-V bulk semiconductor proportional to $p^3$ (the bulk Dresselhaus constant) and $a_{SO}$ is the constant determined by the parameters of the band structure and by the magnitude of the spin-orbit interaction. Averaging is performed over the envelope functions of the ground two-dimensional subband. Interfacial contributions to the Dresselhaus and the Rashba constants are not presented. One of the aims of this work is to derive them. It is noteworthy that they vanish within the method of smooth envelope functions for the model of a high heterobarrier considered below.

In Eqs. (\ref {const BIA SIA}), it is significantly used that the envelope function method is applicable throughout the entire space, including the heteroboundary region \cite{s8, s9}. However, this method is applicable for the description of only smooth (at atomic scales) fields and is inapplicable for the real case of atomically sharp interfaces. Information on the microscopic structure of the heteroboundary can be taken into account in the corresponding boundary conditions for the envelope function.

The problem of such boundary conditions has a long history. Theoretical works concerning this problem can be conditionally classified into two groups. The
most numerous works devoted to the derivation of ''two-sides'' boundary conditions relating the envelope functions and their derivatives on the left and
right of the interface belong to the first group. They involve various approaches to the solution of mathematical problems associated, in particular, with the possible singular behavior of the envelope functions at the  heteroboundary \cite {s5,s9, s10, s11, s12, s13, s14}.

 The works of the second group are devoted to the derivation of ''one-sides'' boundary conditions at the (crystal-high barrier) interface (in particular, at the crystal-vacuum interface). Such problems appear, e.g., in the description of surface (interface) Tamm-type states. This work belongs to a few works of the second group.

We briefly describe some results obtained in the works of the second group. Without accounting of spin, the microscopic derivation of the boundary conditions for the envelope functions at the stepwise interface between a semiconductor ($z > 0$) and vacuum ($z < 0$) evidently was reported for the first time in \cite{s15, s16}. The boundary conditions obtained in those works contain boundary parameters analytically (but complicatedly) expressed in terms of the total band structure of a semiconductor analytically continued to the region of complex quasimomenta. The numerical determination of these parameters is an unsolved problem. In the single-band limit, the boundary condition is a linear relation
between a function and its derivative with the single boundary parameter $R$ with the dimension of length. Physically, this parameter represents the localization depth of a shallow Tamm state when it exists (under the condition $R > 0$). The much simpler derivation of this boundary condition from the  condition that the effective Hamiltonian for envelope functions on the half-space bounded by an impenetrable barrier is Hermitian was given in \cite{s17}. In this phenomenological approach, the parameter $R$ should be determined from an experiment. The high-barrier model is applicable when the
interface length $R$ significantly exceeds the penetration length under the barrier. The effect of the spin-orbit interaction on the boundary condition for envelope functions, as well as the spin splitting of two-dimensional states in the conduction band of a semiconductor with bulk inversion symmetry, was analyzed in \cite{s18} within the generalization of the approach used in \cite{s17}. The nonparabolic generalization of the boundary condition obtained in \cite{s18} and the Bychkov-Rashba term in an asymmetric quantum well with infinite barriers were presented in \cite{s19}.

 In this work, we analyze the effect of the atomically sharp heteroboundary on the effective two-dimensional Hamiltonian and the spin splitting of the spectrum of two-dimensional electrons in crystals with bulk inversion asymmetry. The band discintinuity at the heteroboundary is assumed to be large and the heterobarrier is considered as impenetrable. The heterobarrier is characterized by a certain boundary condition for the envelope functions. In the absence of magnetic field, this leads to the renormalization of the expressions for the constants $\alpha_{BIA} $ and $\alpha _{SIA} $. We also study the spin (Zeeman) splitting of the energy of electrons in the oblique magnetic field $\boldsymbol B$  having the quantizing $B_z$ component.

The Zeeman splitting value is usually a linear function of the magnetic field with the proportionality coefficient equal to the Bohr magneton $\mu_B$ multiplied by the $g$-factor. The $g$-factor of electron in crystal $g^*$ differs from the value $g_0=2$ in vacuum because of the spin-orbit interaction, and depends strongly on the band structure \cite{s20}. However, it is still isotropic in cubic crystals. In heterostructures with a symmetric quantum well grown in the direction $z || [001]$, the components of the $g$-factor tensor along and across the well become different \cite{s21}. The effect is explained by the nonparabolicity of the conduction band \cite{s22}. The nonparabolicity effect will be neglected below. Nonzero off-diagonal (in cubic axes) components of the $g$-factor tensor appear in heterostructures with an asymmetric  quantum well \cite{s23}.

The dependence of the $g$-factor on the quantizing component of the magnetic field $B_z$ and number of the Landau level $N$  was revealed in the recent high-precision measurements of spin resonance in GaAs quantum wells in the quantum Hall effect regime \cite{s24,s25}. This unusual behavior of $g(B_z)$ motivates the formulation of the problem in this work.

The phenomenological boundary condition for the envelope functions in the conduction band is derived in Section 2 using the Hermiticity of the effective multiband Hamiltonian on the half-space and the time reversal invariance.

In Section 3, the problem involving the simple single-band Hamiltonian and a complex boundary condition is unitarily transformed to a simpler problem with the renormalized Hamiltonian and standard (zero) boundary condition. The further averaging over the fast motion along the quantum confinement axis yields (at $B = 0$) the effective two-dimensional Hamiltonian given by Eq. (\ref{2D ham}) with the constants  $\alpha_{BIA} $ and $\alpha _{SIA} $ containing interfacial contributions.

The $g$-factor of two-dimensional electrons is calculated in Section 4. A similar transition to the renormalized two-dimensional Hamiltonian of the conduction band is performed. The components of the $g$-factor tensor are found after averaging over the $N$th Landau level. 

The results are compared to the experiment reported in \cite{s25} in Section 5 and are discussed in Section 6.

\section {Boundary condition for the envelope functions of a conduction band electron.} 

We consider a one-side doped heterojunction (001) GaAs / Al$_x$Ga$_{1-x}$As. Electrons occupy the region  $z \ge 0$ and move in a well  created by the atomically smooth potential $V(z)$ at $z > 0$ and a sharp impenetrable barrier at $z = 0$. We introduce boundary conditions for the envelope functions at $z=0$.

In the framework of the multiband envelope function method, the dynamics of a conduction electron at $z > 0$ is described by the Kohn-Luttinger equation

\begin{multline}
\label{multiband ham}
\Biggl[ \left(E_n(0)+V(z)\right) \delta_{nn'}+\frac{{\bf \hat p p}_{nn'}}{m_0}+\\+\frac{\hbar}{4 m_0^2 c^2}({\bf p} [\bm{\sigma}\times \bm{ \nabla} V_0])_{nn'} \biggr] \Phi _{n'}=E \Phi _n, 
\end{multline}
where $n$ is the  number of the band, $E_n(0)$ is the energy of the extremum in the $n$th band,  $\Phi _n$ is the set of the envelope functions, $\boldsymbol p$ is the momentum operator, $\boldsymbol p_{nn'}$ is the matrix element of the momentum operator on Bloch functions of the center of the Brillouin zone, $m_0$ is the mass of the free electron, and  $\hbar ({\bf p} [\bm{\sigma}\times \bm{\nabla} V_0])_{nn'}/4 m_0^2 c^2$ is the matrix element of the spin-orbit interaction on Bloch functions.

The requirement of the Hermiticity of Hamiltonian (\ref {multiband ham}) on the half-space after integration by parts reduces to the vanishing of the surface contribution.
This is equivalent to the vanishing of the normal component of the current operator at the boundary:
\begin{equation}
\label {ermit general}
( \Phi _{\lambda}^{\dagger} \hat {\bf v}_z \Phi _{\nu} )|_{z=0} = 0,
\end{equation}
where $\hat {\bf v}_z$ is the off-diagonal velocity matrix 
$$({\bf v}_z)_{nn'} =\partial _{p_z}(H_{nn'}).$$

To describe the band structure of III-V compounds with a moderately wide band gap $E_g$, the Kane model \cite{s26} is usually used. Four bands (eight bands taking spin into account) are considered: conduction band, heavy hole band, light hole band, and split-off band. However, this model disregards the
lack of inversion symmetry in the crystal potential. For this reason, we use the 14 band extended Kane model \cite{s3, s5, s27, s28, s29}. In this case, in addition to the $\Gamma _{6c}$, $\Gamma _{8v}$ and $\Gamma_{7v}$ symmetry bands included in the standard Kane model, higher $\Gamma _{8c}$ è  $\Gamma _{7c}$ bands are taken into account. Hamiltonian (\ref {multiband ham}) is a $14 \times 14$ matrix. In this case, there are three nonzero matrix elements of the momentum operator: $P_0$ (between the functions of the  $\Gamma _{6c}$ band and the functions of the  $\Gamma _{7v}$ and  $\Gamma _{8v}$ bands), $P_1$ (between the functions of the $\Gamma _{6c}$ band and the functions of the  $\Gamma _{7c}$ and $\Gamma _{8c}$ bands), and $Q$ (between the functions of the $\Gamma _{7v}$ and  $\Gamma _{8v}$ bands and the functions of the $\Gamma _{7c}$ and $\Gamma _{8c}$ bands). The nonzero value of $P_1$ is due to the absence of the inversion center in the III-V crystal. The matrix element $\Delta^-$ associated with the spin-orbit interaction between the $\Gamma _{7v}$ and $\Gamma _{8v}$ bands and the $\Gamma _{7c}$ and $\Gamma _{8c}$ bands is also nonzero. The problem becomes
$$\hat H_{14 \times 14}\Phi =E\Phi$$
with the general constraint given by Eq. (\ref {ermit general}). Since only the spinor corresponding to the conduction band $\Gamma _{6c}$ is large in the multicomponent function $\Phi$, we make the unitary transformation \cite{s3} $\Phi=e^S \phi$ (taking into account the $kp$ terms up to the third order inclusively) which reduces the Hamiltonian to a single-band Hamiltonian with the effective mass $m^*$.

The resulting three-dimensional Hamiltonian of the conduction band contains the contributions $\hat H_{BIA}$ and $\hat H_{SIA}$ describing the spin splitting due to the crystal inversion asymmetry and the asymmetry of the well:

\begin{equation}
\label{ham}
\hat H=\frac{{\hat p}^2}{2m^*}+V(z)+\hat H_{BIA}+\hat H_{SIA}.
\end{equation}
\begin{multline}
\label{ham dress}
\hat H_{BIA}=\frac{\gamma_c}{\hbar^3 } \biggl[\sigma_x p_x (p_y^2-\hat p_z^2)+\sigma_y p_y (\hat p_z^2-p_x^2)+\\+\sigma_z \hat p_z (p_x^2-p_y^2)\biggr],
\end{multline}
\begin{equation}
\label{ham rashba}
\hat H_{SIA}=a_{SO}(\sigma_x p_y - \sigma_y p_x)\partial_z V(z).
\end{equation}

The same transformation reduces Eq. (\ref {ermit general}) to the following constraint for the spinor  $\phi=(\phi_1,\phi_2)^t$:
\begin{equation}
\label {ermit ext kane}
\left. \left(\phi_{\lambda}^{\dagger }  \tilde{v}_z  \phi_{\nu} + (\tilde{v}_z \phi_{\lambda})^{\dagger}  \phi_{\nu} \right) \right|_{z=0} = 0,
\end{equation}
\begin{multline*}
\tilde{v}_z=\frac{ \hat p_z}{m^*} + \frac{i \chi}{m^*}(\sigma_x p_y -\sigma_y p_x)+\frac{2 \gamma_c}{{\hbar^3}} (\sigma_y p_y -\sigma_x p_x) \hat p_z +\\+ \frac{ \gamma_c}{{\hbar^3}} \sigma_z (p_x^2 - p_y^2)+\frac{ib}{m^*} \hbar  \partial_z V(z),
\end{multline*}
Here,  $b \propto 1/E_g$, $\chi=(g_0 - g^*)m^*/2m_0$ are the bulk parameters ($\chi \simeq 0.082$ for GaAs). The analytical calculations were performed using the Wolfram Mathematica software.

Constraint given by Eq. (\ref {ermit ext kane}) is insufficient for the determination of the boundary conditions. Following \cite{s30}, we require the invariance of Eq. (\ref {ermit ext kane}) under time reversal $\hat T = i \sigma_y \hat K$ , where $\hat K$ is the complex conjugation operator. As a result, we obtain $T$ invariant boundary conditions

\begin{multline}
\label {GU ext kane}
\biggl [1 + i \frac{R \hat p_z}{\hbar}+ \frac{\chi R}{\hbar} (\sigma_y p_x -\sigma_x p_y) + i \frac{2m^* \gamma_c R}{{\hbar^4}} (\sigma_y p_y -\sigma_x p_x) \hat p_z +\\+ i \frac{m^* \gamma_c R}{{\hbar^4}} \sigma_z (p_x^2 - p_y^2)\biggr] \phi \Biggr|_{z=0}=0.
\end{multline}
The second and third terms in boundary condition (\ref {GU ext kane}) were previously known (see \cite{s17,s18}, respectively), whereas the last two terms are new and are due to the spin-orbit interaction at the interface and to the lack of inversion symmetry in the bulk crystal. These terms can be obtained immediately from the Hermiticity of the single-band Hamiltonian specified by Eqs. (\ref{ham}) -- (\ref{ham rashba}). For this reason, boundary condition (\ref {GU ext kane}) is not restricted by the extended Kane model. The
real parameter $R$ depends on the microscopic structure of the boundary. Its physical meaning is explained above.

In the framework of the method of smooth envelope functions, zero boundary conditions are standard for the case of the high heterobarrier under consideration. Below, we assume that the difference of boundary condition (\ref {GU ext kane}) from zero one is small. To this end, the length $R$ should be much smaller than the characteristic lengths of the problem, including the thickness of the two-dimensional layer along the $z$ axis. This justifies the use of perturbation theory in the small parameter $R$. We transform boundary condition (\ref {GU ext kane}) to the more convenient form $\hat \Gamma \phi |_{z=0}=0$ with the operator $\hat \Gamma$ , which is unitary including terms up to $R^2 p_i p_z$ ($i = x, y$) inclusively:

\begin{multline}
\label {preobr GU}
\hat \Gamma = 1 + i \biggl[ \frac{R \hat p_z}{\hbar} + \frac{2m^* \gamma_c R}{{\hbar^4}} (\sigma_y p_y -\sigma_x p_x) \hat p_z +\\+\frac{m^* \gamma_c R}{{\hbar^4}} \sigma_z (p_x^2 - p_y^2) + \frac{\chi R^2}{\hbar ^2}(\sigma_x p_y - \sigma_y p_x) \hat p_z \biggr].
\end{multline}

\section{Interfacial contribution to the effective two-dimensional Hamiltonian at zero magnetic field.}

The unitary (with the accuracy indicated above) transformation $\psi=\hat \Gamma \phi$ reduces the problem to the new problem
$$ (\hat H + \delta \hat H) \psi = E \psi, \qquad \psi |_{z=0}=0.$$
The correction to the three-dimensional Hamiltonian has the form
\begin{multline*}
\delta \hat H = R \partial_z V +  \frac{\chi R^2}{\hbar}(\sigma_x p_y - \sigma_y p_x) \partial_z V +\\+ \frac{2m^* \gamma_c R}{{\hbar^3}} (\sigma_y p_y -\sigma_x p_x)\partial_z V.
\end{multline*}
The averaging of $\delta \hat H$ over the fast motion of the electron along the $z$ axis leads to the effective spin two-dimensional Hamiltonian given by Eq. (\ref {2D ham}), besides the energy shift $R(\partial_z V)_{00}$ which is insignificant here. The modified constants $\alpha_{BIA}$ and $\alpha_{SIA}$ contain the contributions depending on the interfacial parameter $R$:

\begin{equation}
\label{Dress const}
\alpha_{BIA} = \frac{2 m^* \gamma_c}{\hbar ^3} \left( \frac{(\hat p_z ^2)_{00}}{2m^*}+e F R \right),
\end{equation}
\begin{equation}
\label{Rashba const}
\alpha_{SIA} = eF \left( a_{SO}+\frac{\chi R^2}{\hbar} \right),
\end{equation}
where $F=(\partial_z V/e)_{00}$ is the average electric field in the
heterostructure and $e$ is the elementary charge. Formulas (\ref{Dress const}) and (\ref{Rashba const}) constitute one of the main results of the work.

\section{Interfacial contribution to the zeeman splitting of the landau levels.}

We now analyze the effect of the magnetic field on the spin splitting of the electronic spectrum. We replace the momentum operator by the generalized
momentum operator \cite{s31}:  $\hat p_i \rightarrow \hat  \pi _i = \hat p_i + eA_i/c, $ where $\boldsymbol A$  is the vector potential of the magnetic field. The noncommuting components of the momentum operators should be replaced by the symmetrized combinations: $\hat {\pi}_i \hat {\pi}_j \rightarrow (\hat {\pi}_i \hat { \pi}_j +\hat {\pi}_j \hat {\pi}_i)/2 \equiv \{\pi_i,\pi_j\}$.

Applying those changes to the Hamiltonian specified by Eqs. (\ref {ham}) -- (\ref {ham rashba}) and boundary conditions given by Eqs. (\ref{GU ext kane}), (\ref {preobr GU}), we perform the unitary transformation of wavefunctions as in Section 3 and pass to a new problem with the transformed Hamiltonian and zero boundary condition. The correction to the three-dimensional Hamiltonian contains two contributions, orbital and spin:

\begin {equation}
\delta \hat H=\delta \hat H_0+\delta \hat H_s,
\end {equation} 
where
\begin {equation}
\label{51}
\delta \hat H_0= -\frac{eR}{m^* c} \hat \pi_x B_y + \frac{eR}{m^* c} \hat \pi_y B_x + R \partial _z V,
\end {equation}

\begin{multline}
\delta \hat H_s= - \frac{qR\gamma _c}{\hbar ^3c} \biggl(\sigma_x(\pi_y^2B_y-\pi_z^2B_y-2\{\pi_x \pi_y\}B_x)+\\+\sigma_y(\pi_x^2B_x-\pi_z^2B_x-2\{\pi_x \pi_y\}B_y)+\sigma_z(2\{\pi_x \pi_z\}B_y+2\{\pi_y \pi_z\}B_x)\biggr)+\\+\frac{a_{SO}qR\partial _z V}{c}(\sigma_xB_x+\sigma_yB_y)+ \frac{2m^* \gamma_c R}{{\hbar^3}} (\sigma_y \hat \pi_y -\sigma_x \hat \pi_x)\partial _z V -\\- \frac{qR\gamma _c}{\hbar ^3c} \biggl(\sigma_y(-2\pi_y^2B_x+2\pi_z^2B_x+2\{\pi_x \pi_y\}B_y-2\{\pi_x \pi_z\}B_z)-\\-\sigma_x(2\pi_x^2B_y-2\pi_z^2B_y-2\{\pi_x \pi_y\}B_x+2\{\pi_y \pi_z\}B_z)\biggr) +\\+ \frac{2 m^* \gamma_c R\mu_B g^*}{\hbar ^4} \biggl(-\sigma_z  B_x\{\pi_y \pi_z\} + \sigma_x B_z\{\pi_y \pi_z\} - \sigma_z B_y\{\pi_x \pi_z\}+\\ +\sigma_y B_z\{\pi_x \pi_z\} \biggr)+\frac{\mu_B g^* m^* \gamma_c R}{\hbar ^4} (\hat \pi _x ^2 - \hat \pi _y ^2)(\sigma _ y B_x - \sigma_x B_y) +\\+ \frac{2qR\gamma _c}{\hbar ^3c}\sigma_z (2B_z\{\pi_x \pi_y\}-B_y\{\pi_x \pi_z\}-B_x\{\pi_y \pi_z\})+\\+\frac{ \chi R^2}{\hbar} (\sigma_x \hat \pi_y -\sigma_y \hat \pi_x)\partial _z V - \frac{\chi R^2q}{2\hbar m^*c} \biggl(\sigma_x(-2\pi_y^2B_x+2\pi_z^2B_x+\\+2\{\pi_x \pi_y\}B_y-2\{\pi_x \pi_z\}B_z)-\sigma_y(2\pi_x^2B_y-2\pi_z^2B_y-2\{\pi_x \pi_y\}B_x+\\+2\{\pi_y \pi_z\}B_z)\biggr) + \frac{ \chi R^2\mu_B g^* }{ \hbar ^2} \biggl(\sigma_z B_y\{\pi_y \pi_z\} -\sigma_y B_z \{\pi_y \pi_z\}+\\ + \sigma_z B_x\{\pi_x \pi_z\} - \sigma_x B_z\{\pi_x \pi_z\}\biggr)
\end{multline}

After averaging over the envelope function of the lowest subband, we obtain a rather lengthy expression for the effective two-dimensional Hamiltonian.
Finally, we perform averaging over the eigenfunctions of the orbital part of the Hamiltonian including contribution (\ref{51}):

\begin{multline}
\label{spin ham}
\hat H_N=\frac{\hbar e|B_z|}{m^*c}\left(N+\frac{1}{2}\right)+\frac{q^2}{2m^*c^2}(B_x^2+B_y^2)((z^2)_{nn}-z_{nn}^2)-\\-\frac{e^2R^2}{2m^* c^2}(B_x^2+B_y^2)+eFR+\frac{|\mu_B|}{2} g_{ij}(|B_z|) \sigma_i B_j.
\end{multline}
Spin Hamiltonian (\ref{spin ham}) describes the Zeeman splitting of the $N$th Landau level ($N$ = 0, 1, 2,...) shifted in energy owing to the magnetic $(B_x, B_y)$ and electric ($F$)
fields. The tensor $g_{ij}(B)=g_{ij}(0)+d_{ij}|B_z|$ is anisotropic in the $(i, j) = (x, y)$ plane and nonanalytically depends on $B_z$ because of the specificity of the Landau quantization.

The interface contributions to the diagonal components of the $g$-factor depend on the length $R$ and the bulk spin-orbit constant $\chi$:

\begin{equation}
\label{diag}
 \delta g_{xx}^{int}=\delta g_{yy}^{int}=\frac{4\chi R^2m_0}{\hbar^2}\frac{(p_z^2)_{00}}{m^*}.
\end{equation}

\begin{equation}
\label {derevativ diag g}
d_{xx}=d_{yy}=-\frac{4\chi R^2 m_0e}{m^*c\hbar}(N+\frac{1}{2}),
\end{equation}

The off-diagonal components also include the interface ($R$) and bulk ($\gamma_c$) contributions :
\begin{multline}
\label{zero g}
g_{xy}(0)=g_{yx}(0)=\frac{4m_0\gamma_c}{\hbar ^4}  ( (p^2_z)_{00} z_{00}-(p^2_z z)_{00})+\\+\frac{8m_0\gamma_cR}{h^4}(p_z^2)_{00},
\end{multline}

\begin{equation}
\label{derevativ g}
d_{xy}=d_{yx}=-\frac{8\gamma_cRm_0e}{\hbar^3 c}(N+\frac{1}{2}).
\end{equation}
The first term in Eq. (\ref {zero g}) was obtained in \cite{s23}. The second term is the sought-for interfacial contribution. Furthermore, the off-diagonal components of the $g$-factor are linear functions of the quantizing component of the magnetic field $|B_z|$. The proportionality coefficient given by Eq. (\ref {derevativ g}) depends linearly on the number of the corresponding Landau level and is determined only by the interfacial contribution.

\section{Comparison with the experiment.}

We compare our results with the experimental data reported in \cite{s25}, which were obtained with an asymmetrically doped 20-nm-wide  GaAs/Al$_{0.3}$Ga$_{0.7}$As quantum well with the electron 
density  $n_s=4.4\times 10^{11}$ sm$^{-2}$ ($F=0.304 \times 10^5$ V/sm).

The off-diagonal components of the $g$-factor tensor are equal to each other, as in \cite{s23}. Consequently, in the axes coinciding with the $[110]$, $[1\bar {1}0]$ and $[001]$ directions, the $g$-factor tensor is diagonal in agreement with \cite{s24, s25}.

In our notations, the data reported in \cite{s25} can be represented as $g_{x'x'}(0)= -0.292$, $g_{y'y'}(0)=-0.347$, $d_{x'x'}=0.002$ T$^{-1}$, $d_{y'y'}=0.012$ T$^{-1}$. The corresponding components of the $g$-factor tensor in the cubic axes are  $g_{xx}(0)=-0.3195$, $g_{xy}(0)=0.0275$, $d_{xy}=-0.005$ T$^{-1}$. The difference of the diagonal components from the bulk value $g^* = -0.44$ is associated with the interfecial contribution given by Eq. (\ref{diag}) and with the nonparabolicity contribution  $\delta g^{np}$ omitted above. In this approximation two feasible values of $R$ ($R_1 = 10$ \r{A} and $R_2 = -10$ \r{A})  can be obtained from Eq. (\ref{diag}). Accordingly, comparison of Eq. (\ref{zero g}) with the experimental data yields $\gamma_c^{(1)}=4$ eV $\times$ \AA $^3$ and $\gamma_c^{(2)}=14$ eV $\times$ \AA $^3$. 

We now consider the dependence $g_{xy}(|B_z|)$, which is more subtle effect. From (\ref{derevativ g}) we obtain $d_{xy}^{(1)}=-0.001$ T$^{-1}$ and $d_{xy}^{(2)}=0.003$ T$^{-1}$. The values of $\gamma_c$ is smaller than the values reported in \cite{s32}. A possible reason for these discrepancy is the disregard of the nonparabolic contribution.

The smallness of $R$ as compared to the quantum confinement lengths justifies the use of perturbation theory:  $z_{00}=89$ \AA; the average value of  $R\partial_z$ is 0.003.

\section{Discussion.}

Boundary condition (\ref{GU ext kane}) describes  the atomically sharp heterointerface of the GaAs/AlGaAs type with a large discontinuity of the conduction band, the bulk inversion asymmetry, and the spin-orbit interaction in the bulk and at the interface.

The spin splitting of electron Landau levels is anisotropic, nonlinear, and nonanalytic as a function of the quantizing component of the magnetic field.
The interface contributions to the diagonal and off-diagonal components of the $g$-factor tensor are obtained. The values of $R$ and $\gamma_c$ were determined by comparison with the experiment \cite{s25}. 

It is worth noting that not only the Rashba constant $\alpha _{SIA}$ but also the Dresselhaus constant $\alpha _{BIA}$ (more slightly) depends on the electric field $F$ "pressing" electrons to the interface. This dependence makes it possible to control the indicated parameters. For the parameters $F=0.304 \times 10^5$ V/sm, $\gamma_c^{(1)}=4$ eV$\times$ \AA$^3$, $R_1=10$\AA, we obtain the following renormalized constants: $\alpha_{BIA}\times\hbar=1.1$  meV$\times$\r{A}  instead of $\alpha_{BIA}^{0}\times\hbar=0.9$  meV$\times$\r{A}; $\alpha_{SIA}\times\hbar=4$ meV$\times$\r{A}~ instead of $\alpha_{SIA}^{0}\times\hbar=1.4$  meV$\times$\r{A}. For the parameters $F=0.304 \times 10^5$ V/sm, $\gamma_c^{(2)}=14$ eV$\times$ \AA$^3$, $R_2=-10$\AA, we obtain: $\alpha_{BIA}\times\hbar=2.6$  meV$\times$\r{A}  instead of $\alpha_{BIA}^{0}\times\hbar=3.4$  meV$\times$\r{A}; $\alpha_{SIA}\times\hbar=4$ meV$\times$\r{A}~ instead of $\alpha_{SIA}^{0}\times\hbar=1.4$  meV$\times$\r{A}. The main contribution to $\alpha_{SIA}$ comes from the interface. Particular values can vary with the inclusion of the nonparabolicity effect.

The above results are strictly valid only in the limit of impenetrable heterobarrier. It is important that the envelope functions in this approximation are discontinuous at the interface because of the nonperturbative
effect of the interface potential. For this reason, the approach used in this work allows to describe, for instance, shallow Tamm states even in the single-band approximation \cite{s15,s16,s17}.

The continuity of single-band envelope functions at the heterointerface is most often used in studies. The theory is usually developed with two-sided
boundary conditions, when there is penetration under the barrier \cite{s3,s4,s5,s9,s32}, and cannot describe Tamm states. For this reason, it is difficult to directly compare our results with the known data. In particular, the interfacial spin contributions considered using the envelope function method throughout the entire space in \cite{s9} (see also Eq. (2.120) in \cite{s5}) disappear when functions under barrier formally vanish. In \cite{s33}, the specific microscopic mechanism associated with the mixing of the envelope functions of light and heavy holes at the atomically sharp heterointerface \cite{s33} was generalized to the conduction band. The corresponding spin contribution to the three-dimensional Hamiltonian has the Dresselhaus-type structure, but is singular in the coordinate space. The averaging of this singularity also results in the disappearance of the interfacial contribution in the limit of the impenetrable barrier.

At the same time, real heterobarriers always have finite heights. For example, a underbarrier length of 13 \r{A} for the GaAs/Al$_{0.3}$Ga$_{0.7}$As is the order of the parameter $|R| = 10$ \r{A} of our theory. Even weak penetration under the barrier, where the sign of the $g$-factor changes, can noticeably affect the parameters extracted from the experiment \cite{s22}. This problem requires a special consideration.

We are grateful to I.V. Kukushkin for detailed discussions of the experimental results stimulating the formulation of this problem and to E.L. Ivchenko, M.M. Glazov, and A.V. Shchepetilnikov for useful remarks. This work was supported in part by the Russian Foundation for Basic Research, project no.  ¹ 11-02-01290.


\begin{thebibliography}{10}

\bibitem{s1}
F. Malcher, G. Lommer, U. Rossler, Superlatt. Microstruct. {\bf 2}, 267 (1986).

\bibitem{s2}
M. I. D'yakonov and V. Yu. Kocharovskii, Sov. Phys.
Semicond.{\bf 20}, 178 (1986).

\bibitem{s3}
R. Winkler, Spin-orbit coupling effects in two-dimensional electron and hole systems, Springer, Berlin, 2003.

\bibitem{s4}
E.\,I. Ivchenko, G.\,E. Pikus, Superlattices and other heterostucture, Springer, Berlin, 1995.

\bibitem{s5}
E.\,L. Ivchenko, Optical Spectroscopy of Semiconductor Nanostructures, Alpha Science, Harrow, UK, 2005.

\bibitem{s6}
Yu. A. Bychkov and E. I. Rashba, JETP Lett. {\bf 39}, 78
(1984).

\bibitem{s7}
E.\,I. Rashba and V.\,I. Sheka, in: Landau Level Spectroscopy. Eds. G. Landwehr and E.\,I. Rashba, North-Holland, Amsterdam, 1991, p. 131.

\bibitem{s8}
L. Leibler, Phys. Rev. B {\bf 16}, 863 (1977).

\bibitem{s9}
W. Zawadzki, P. Pfeffer, Semicond. Sci. Technol. {\bf 19}, R1 (2004).

\bibitem{s10}
B.\,A. Foreman, Phys. Rev. B {\bf 72}, 165345 (2005).

\bibitem{s11}
E. E. Takhtamirov and V. A. Volkov, JETP {\bf 116}, 1843 (1999).

\bibitem{s12}
A.\,V. Rodina, A.\,Yu. Alekseev, A.\,L. Efros, et al., Phys. Rev. B {\bf 65}, 125302 (2002).

\bibitem{s13}
E.\,E. Takhtamirov, V.\,A. Volkov, Semicond. Sci. Technol. {\bf 12}, 77 (1997).

\bibitem{s14}
E. Takhtamirov, R.\,V\,N. Melnik, New J. Phys. {\bf 12}, 123006 (2010).

\bibitem{s15}
V. A. Volkov and T. N. Pinsker, Sov. Phys. JETP {\bf70},
2268 (1976).

\bibitem{s16}
V. A. Volkov and T. N. Pinsker, Sov. Phys. JETP {\bf72}, 1087
(1977).

\bibitem{s17} 
 V.\,A. Volkov,  T.\,N. Pinsker, Surf. Sci. {\bf 81}, 181 (1979).

\bibitem{s18}
F. T. Vas'ko, JETP Lett. 30, 574 (1979).

\bibitem{s19}
A.\,V. Rodina, A.\,Yu. Alekseev, Phys. Rev. {\bf B 73}, 115312 (2006).

\bibitem{s20}
L.\,M. Roth, Phys. Rev. {\bf 118}, 1534 (1960).

\bibitem{s21}
V. K. Kalevich and V. L. Korenev, JETP Lett. {\bf 56}, 253 (1992).

\bibitem{s22}
E. L. Ivchenko and A. A. Kiselev, Sov. Phys. Semicond.
26, 1471 (1992).

\bibitem{s23}
V. K. Kalevich and V. L. Korenev, JETP Lett. 57, 571
(1993).

\bibitem{s24}
 Yu.\,A. Nefyodov, A.\,V Shchepetilnikov, I.\,V. Kukushkin, et al., Phys. Rev. B {\bf 83}, 041307 (2011).

\bibitem{s25}
Yu.\,A. Nefyodov, A.\,V Shchepetilnikov, I.\,V. Kukushkin, et al., Phys. Rev. B {\bf 84}, 233302 (2011).

\bibitem{s26}
E.\,O. Kane, Phys. Chem. Solids. {\bf 1}, 249 (1957).

\bibitem{s27}
U. R\"ossler, Solid State Comm. {\bf 49}, 943 (1984).

\bibitem{s28}
H. Mayer, U. R\"ossler, Phys. Rev. B {\bf 44}, 9048 (1991).

\bibitem{s29}
P. Pfeffer, W. Zawadzki, Phys. Rev. B {\bf 41}, 1561 (1990).

\bibitem{s30}
V. A. Volkov and T. N. Pinsker, Sov. Phys. Solid State
23, 1756 (1981).

\bibitem{s31}
G.\,M. Luttinger, W. Kohn,  Phys. Rev. {\bf 97}, 869 (1955).

\bibitem{s32}
J. Fabian,A. Matos-Abiaguea, C. Ertlera, et al., Acta physica slovaca {\bf 57}, 565 (2007).

\bibitem{s33}
E.\,L. Ivchenko, Y.\,A. Kaminski, U. R\"ossler, Phys. Rev. B {\bf 54}, 5852 (1996).

\bibitem{s34}
 U. R\"ossler, J. Kainz, Solid State Comm. {\bf 121}, 313 (2002).


\end{thebibliography}
\end{document}